\documentclass[twocolumn]{pasj00}

\begin{document}
\SetRunningHead{Fujita et al.}{Outstanding Disk in Hydra A}

\title{Discovery of an Outstanding Disk in the cD Galaxy of the Hydra A
Cluster\thanks{Based on data collected at Subaru Telescope, which is
operated by the National Astronomical Observatory of Japan.}}

\author{Yutaka \textsc{Fujita},\altaffilmark{1}
        Nobuhiro \textsc{Okabe},\altaffilmark{2,3}
	Kosuke \textsc{Sato},\altaffilmark{4}
	Takayuki \textsc{Tamura},\altaffilmark{5}
	Satoki \textsc{Matsushita},\altaffilmark{2}
	Hiroyuki \textsc{Hirashita},\altaffilmark{2}
	Masanori \textsc{Nakamura},\altaffilmark{2}
	Kyoko \textsc{Matsushita},\altaffilmark{4}
	Kazuhiro \textsc{Nakazawa},\altaffilmark{6}
        and
        Motokazu \textsc{Takizawa}\altaffilmark{7}
        }
\altaffiltext{1}{Department of Earth and Space Science, Graduate School
of Science, Osaka University, Toyonaka, Osaka 560-0043}
\altaffiltext{2}{Academia Sinica, Institute of Astronomy and
Astrophysics, P.O. Box 23-141, Taipei 10617, Taiwan}
\altaffiltext{3}{Kavli Institute for the Physics and Mathematics of the
Universe (WPI), Todai Institutes for Advanced Study, University of
Tokyo, 5-1-5 Kashiwanoha, Kashiwa, Chiba 277-8583, Japan}
\altaffiltext{4}{Department of Physics, Tokyo University of Science, 1-3
Kagurazaka, Shinjyuku-ku, Tokyo 162-8601}
\altaffiltext{5}{Institute of Space and Astronautical Science, Japan
Aerospace Exploration Agency, 3-1-1 Yoshinodai, Chuo-ku, Sagamihara,
Kanagawa 252-5210}
\altaffiltext{6}{Department of Physics, The University of Tokyo, 7-3-1
Hongo, Bunkyo-ku, Tokyo 113-0033}
\altaffiltext{7}{Department of Physics, Yamagata University,
Kojirakawa-machi 1-4-12, Yamagata 990-8560}


\maketitle

\begin{abstract}
The central cD galaxy of the Hydra A cluster has one of the most
 powerful active galactic nuclei (AGNs) in the nearby Universe
 ($z\lesssim 0.2$). We report on the discovery of a dust lane in the cD
 galaxy using Subaru telescope. The $i'$-band image shows the existence
 of a dark band of the size of $3.6''\times 0.7''$ ($4\:{\rm kpc}\times
 0.8\:{\rm kpc}$), which appears to be quite similar to the dust lane
 observed in Centaurus~A. The morphology indicates that the cold disk
 that seen as the dust lane is almost edge-on and rotates around the
 AGN. Since the minor axis of the dust lane is nearly parallel to the
 radio jets emerging from the AGN, the disk is probably feeding its gas
 into the central black hole. From the absorption, we estimate the
 hydrogen column density of the lane is $N_{\rm H}=2.0\times
 10^{21}\rm\: cm^{-2}$, and the mass of the disk is $\sim 8\times 10^7\:
 M_\odot$. The column density is consistent with constraints obtained
 from Chandra X-ray observations. The age of the disk is $\gtrsim
 4\times 10^7$~yr. The position angle of the disk and the galaxy's
 photometric axis are misaligned, which may imply that the cold gas in
 the disk is brought via galaxy mergers. Our observations may indicate
 that the supply of cold gas by galaxy mergers is required for the most
 intensive feedback from AGNs.
\end{abstract}

\section{Introduction}

The AGNs in the central cD galaxies of clusters sometimes show violent
activities. Such examples are MS 0735.6$+$7421 \citep{mcn05a},
Hercules~A \citep{nul05b}, and Hydra~A \citep{nul05a}. In these
clusters, cluster-scale shock waves have been observed, which indicates
that the central AGNs ejected enormous energy ($\sim
10^{61}$--$10^{62}$~erg) in a short time ($\lesssim 10^8$~yr). The
intracluster medium (ICM) of these clusters should have been strongly
affected by the AGN activities. It seems that Bondi accretion from hot
gas is not sufficient and fueling of cold gas is required for the most
powerful feedback from AGNs \citep{mcn11a}.

The cD galaxy of the Hydra~A cluster has been studied by many
researchers. It harbors a well known radio source classified as a
Fanaroff-Riley type-I (FR-I) source. A pair of jets is ejected from the
AGN (3C~218; \cite{tay90a}), and the huge amount of cosmic-rays
contained in the jets may be stably heating the cool core of the cluster
\citep{fuj13c}. The mass of the central supermassive black holes (SMBHs)
is estimated to be $\sim 10^9\: M_\odot$ \citep{fuj04c,raf06a}. Although
the galaxy is an elliptical, star formation activities ($\lesssim 1\:
M_\odot\:\rm yr^{-1}$) have been discovered within $\sim 5$~kpc from the
center of the galaxy \citep{mcn95a,car98a,wil04a,hol07a}. This may mean
that some amount of gas is cooling in the galaxy, although the existence
of a massive cooling flow has been denied with the ASCA satellite
\citep{ike97a}. Previous observations have indicated that there is a
disk-like structure in the galaxy. In the $U$-band, the surface
brightness profile is elongated at the galaxy center
\citep{mcn95a}. Observations of optical emission-lines
([O{\footnotesize III}], H$\alpha$, and H$\beta$) show that the galaxy
has a rotating component with a velocity of $\sim 300\rm\: km\: s^{-1}$
\citep{hec85a,mel97a}. H\emissiontype{I} absorption is seen toward the
AGN and can be interpreted as a disk \citep{dwa95a,tay96a}. However, the
morphology of the disk has not been well understood. This is mainly
because the cluster is relatively distant ($z=0.054878$) and the angular
resolution of $<1''$ is required to resolve fine structures.

In this letter, we present a Subaru $i'$-band image of the cD galaxy of
the Hydra A cluster with a superb spatial resolution. We show that the
galaxy has a spectacular dust lane. Throughout this letter, we adopt
cosmological parameters of $\Omega_0=0.3$, $\lambda=0.7$, and
$H_0=70\rm\: km\: s^{-1}\: Mpc^{-1}$.  We use a redshift for the Hydra~A
cluster of 0.054878, which gives a scale of $1.1$~kpc per arcsec. The
angular diameter distance and the luminosity distance to Hydra A are
$220$~Mpc and 245~Mpc, respectively.

\section{Observations and Results}
\label{sec:result}

\begin{figure}
  \begin{center}
    \FigureFile(80mm,80mm){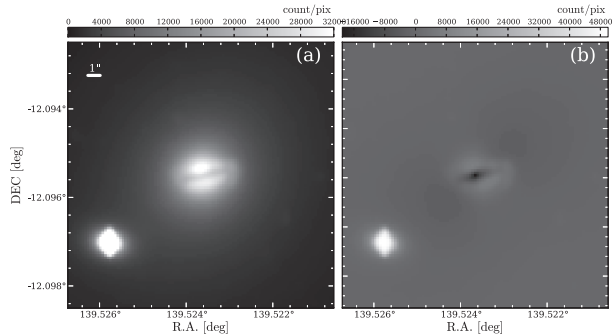}
  \end{center}
  \caption{(a) $i'$-band image of the cD galaxy of
 Hydra~A. (b) Residual of a fit to the galaxy with a Seric profile.}
 \label{fig:img}
\end{figure}

\begin{figure}
  \begin{center}
    \FigureFile(80mm,80mm){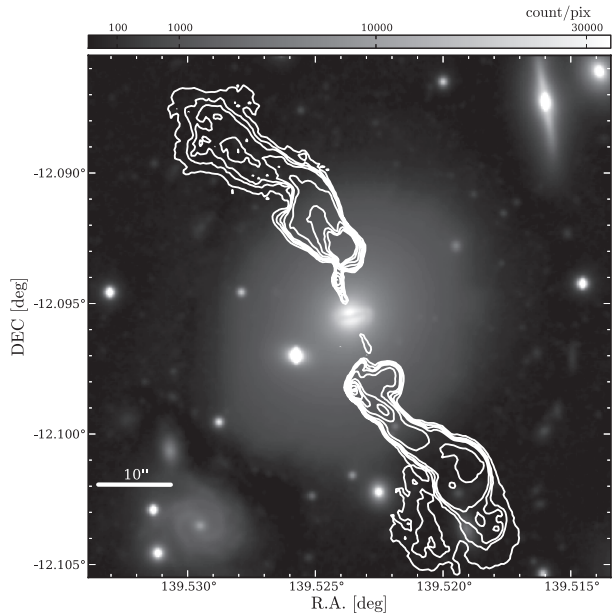}
  \end{center}
  \caption{$i'$-band image of the cD galaxy of Hydra~A (gray). The total
 intensity 6~cm radio map (contours) is superposed
 \citep{tay90a}.}\label{fig:radio}
\end{figure}

We observed Hydra A with the Prime Focus Camera (Suprime-Cam;
\cite{miy02a}) on the Subaru telescope in Hawaii on January 7, 2013. We
use the standard pipeline reduction software for Suprime-Cam, {\sc
SDFRED} \citep{yag02,ouc04}, for flat-fielding, instrumental distortion
correction, differential refraction, PSF matching, and sky subtraction
and stacking. The seeing was $0.65''$. The $i'$-band image of the cD
galaxy is shown in figure~\ref{fig:img}a. A dust lane (dark band) can be
clearly seen in the central region of the cD galaxy.  The size of the
lane is $\sim 3.6''\times 0.7''$ (4~kpc$\times$0.8~kpc).  The position
angle of the minor axis is $\sim 15^\circ$ from the north to the east.
We fit the galaxy image with a Seric profile using {\sc galfit}
\citep{pen10}, taking PSF size into account. The residual from the
best-fits is shown in figure~\ref{fig:img}b. The position angle of the
minor axis of the Seric profile is $40^\circ$, which is close to the
result by \citet{mcn95a} for the $I$-band image ($\sim 55^\circ$), but
is not parallel to the minor axis of the disk ($\sim 15^\circ$).  This
may indicate that the disk formed after the overall structures of the cD
galaxy had formed. The extinction is most effective at the center
(40\%), and it is $A_{\rm I}=0.55$. Since the extinction curves of
elliptical galaxies are not much different from the one for the
Milky-Way \citep{pat07a}, we adopt the Galactic conversion factor
($A_{\rm V}/A_{\rm I}=2.1$; \cite{fit99a}) and obtain $A_{\rm V}=1.1$
and the column density of $N_{\rm H}=2.0\times 10^{21}\rm\: cm^{-2}$
from $N_{\rm H}/A_{\rm V}=1.79\times 10^{21}\rm\: cm^{-2}$
\citep{pre95a}. The value of $N_{\rm H}$ is almost the same as
H{\footnotesize I} absorption ($N_{\rm HI}=1.97\times 10^{21}\rm\:
cm^{-2}$) obtained by \citet{dwa95a}.  If the gas is distributed as a
uniform disk with a radius of $R\sim 2$~kpc and a height of $H\sim
0.8$~kpc, the disk mass is $M_{\rm disk}=8\times 10^7\:
M_\odot$. Figure~\ref{fig:radio} shows the relation of the galaxy to the
radio source. A pair of jets emerges with a position angle of $\sim
25^\circ$ and is almost parallel to the minor axis of the dust lane. The
jet-dust lane configuration is quite similar to that of
Centaurus~A\footnote{http://www.eso.org/public/images/eso0903a/}.

\section{Discussion}

\subsection{X-ray Absorption}

\begin{figure}
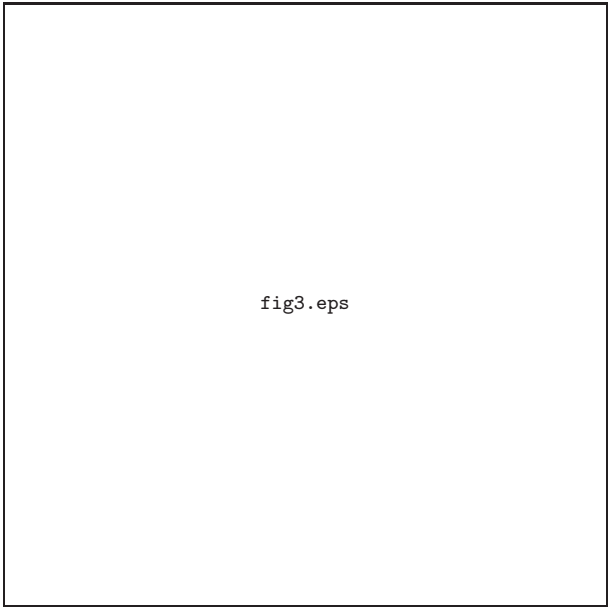

  \begin{center}
    \FigureFile(80mm,80mm){fig3.eps}
  \end{center}
  \caption{Chandra 0.5--5.0 keV image of the cD galaxy of Hydra~A.  The
mosaic image from 200~ksec ACIS-S observations is
exposure-corrected. AGN region ``N'' is a circle with a radius of $1''$,
and the disk region ``D'' is a rectangle of $4.5'' \times 0.5''$,
respectively.}\label{fig:Ximg}
\end{figure}

We have checked X-ray data obtained with the Chandra X-ray
telescope. Hydra A has been observed with Chandra for total 250~ksec.
Figure~\ref{fig:Ximg} shows the exposure-corrected X-ray image of
Chandra ACIS-S for 200~ksec data in the 0.5--5~keV band.  While the
central AGN is prominent, the disk-structure is not seen in the figure.
X-ray spectral analysis for Chandra ACIS-S and ACIS-I data of the total
of 250~ksec was performed to constrain an absorption by the disk with
response files appropriate for each observation in the 0.4--7.0 keV
band.  

We study the spectrum of a $4.5'' \times 0.5''$ region of the disk
(region ``D'' in figure~\ref{fig:Ximg}). We exclude the nuclear region
``N'', because the emission from it can be affected by the absorption in
the vicinity of the SMBH. In fact, we find that the absorption
associated with the AGN is $N_{\rm H} = 2.7^{+0.7}_{-0.6} \times
10^{22}\rm\: cm^{-2}$, which is consistent with the previous result
($N_{\rm H} = 2.8^{+3.0}_{-1.4} \times 10^{22}\rm\: cm^{-2}$
\citep{sam00a}. Owing to the finite size of pixels ($0.5''$), the pixels
we used actually cover the optical disk ($\sim 3.6''\times
0.7''$). First, we extract a spectrum within a $3''$ circle centered on
the AGN excluding the AGN and disk regions (``N'' and ``D'') in order to
estimate the fore- and/or background emissions. The spectrum is
represented by two-temperature (cD galaxy + ICM) thermal models (apec)
with a common metal abundance multiplied by absorption (phabs): ${\rm
phabs}\times({\rm apec}_{\rm cD}+{\rm apec}_{\rm ICM})$. The results are
$kT=1.0^{+0.3}_{-0.2}$ and $3.5^{+0.3}_{-0.3}$ keV with 0.7$\pm 0.2$
solar abundance, and $N_{\rm H}=3.3^{+1.2}_{-1.1} \times
10^{20}$~cm$^{-2}$ (errors of 90\% confidence).  The absorption value is
almost consistent with the Galactic absorption ($N_{\rm H}=4.68 \times
10^{20}$~cm$^{-2}$, \cite{kalberla05}) in the direction of Hydra
A. Fixing these fore- and/or background parameters, we investigate the
upper limit of the absorption by the disk. Here, we assume that the
emission from the disk region is represented by two-temperature thermal
models (cD galaxy + ICM) and only the cooler component (cD galaxy) is
absorbed by the disk: ${\rm phabs}\times({\rm phabs}_{\rm
disk}\times{\rm apec}_{\rm cD}+{\rm apec}_{\rm ICM})$. We employ an
increment of $\delta \chi^2 =2.706$ as the measure for the 90\% upper
limit of the absorption. We find that the upper limit is $N_{\rm H}=3.1
\times 10^{21}$~cm$^{-2}$, which is consistent with the disk absorption
of $N_{\rm H}=2.0\times 10^{21}\rm\: cm^{-2}$
(section~\ref{sec:result}). Note that the absorption by the dust lane on
the both sides of the AGN should be smaller than that in the direction
of the AGN (region ``N'').

\subsection{Nature of the Disk}

Optical emission-lines have been discovered from the cD galaxy of
Hydra~A. If the rotation of the emission-line component corresponds to
that of the dust lane, the rotation velocity is $v_{\rm rot}\sim
300\rm\: km\: s^{-1}$ \citep{hec85a,mel97a}, which gives the rotation
period of $T_{\rm rot}=2\pi R/v_{\rm rot}\sim 4\times 10^7$~yr. The age
of the cold disk should be larger than this period. From the
observations of a shock wave in the ICM, \citet{nul05a} indicated that
there was a strong outburst of the AGN with an energy of $\sim
10^{61}$~erg $\sim 1.4\times 10^8$~yrs ago. This outburst may be related
to the formation of the disk. Note that if all the disk gas at present
($M_{\rm disk}=8\times 10^7\: M_\odot$) flowed into the SMBH, energy
comparable to the previous outburst might be released ($\sim 0.1 M_{\rm
disk}c^2\sim 10^{61}\rm\: erg$).


The angular distance between the X-ray AGN and the center line of the
dust lane is $\Delta\theta\lesssim 0.5''$ or $dl\lesssim 0.55$~kpc
(figure~\ref{fig:Ximg}). If the AGN is located at the center of the
disk, the inclination angle of the disk is $dl/R\lesssim 16^\circ$, and
the disk is almost edge-on. This may also mean that the jets from the
AGN are almost on the plane of the sky at present. On the other hand,
the positions of pairs of outer lobes show that the jets were not being
injected on the plane of the sky and were oriented $\sim 40^\circ$ about
the plane in the past \citep{wis07a}. The change of the jet direction
may be caused by precession \citep{tay90a}.

\subsection{Origin of the Dust Lane}

The cD galaxy is a harsh environment for dust. In the presence of the
hot ICM of $\sim 10^7$~K, collision with hot particles destroys the
dust. The destruction time is
\begin{equation}
 \tau_d=2.4\times 10^5\left(\frac{n_e}{\rm cm^{-3}}\right)^{-1}
\left(\frac{a}{0.1\:\mu\rm m}\right){\rm yr}\:,
\end{equation}
where $n_e$ is the ICM density and $a$ is the radius of dust grains
\citep{dra79a,pat07a}. If we adopt $n_e=0.05$ \citep{nul05a} and
$a=0.1\:\mu\rm m$, the destruction time is $\tau_d=4.8\times 10^6$~yr.
The mass loss rate from evolved stars in the galaxy is
\begin{equation}
 \dot{M}_*\approx 6\times 10^{-6}\left(\frac{D}{\rm Mpc}\right)^2
\left(\frac{S_{12}-0.042\: S_{100}}{\rm mJy}\right)
\: M_\odot \:{\rm yr^{-1}}\:,
\end{equation}
where $D$ is the distance to the galaxy, and $S_x$ is the flux at
$x\:\mu\rm m$ \citep{kan92a}.  For Hydra A, the luminosity distance is
$D=245$~Mpc, and the flux is $S_{12}=5\rm\: mJy$ \citep{don11a}. Since
we are interested in the upper limit of the dust mass $M_d$ and since we
avoid the complications caused by the difference of apertures at
different wave lengths, we assume that $S_{\rm 100}=0$. For these
values, we obtain $\dot{M}_*=2\: M_\odot {\rm yr^{-1}}$. Assuming that the
gas-to-dust ratio is $f=100$, the rate of accumulation of dust is
\begin{equation}
 \frac{dM_d(t)}{dt}=\frac{\dot{M}_*}{f} - \frac{M_d(t)}{\tau_d}
\end{equation}
\citep{pat07a}. This equation can be solved easily; $M_d$ approaches its
final value $M_{d0}=9.5\times 10^4\: M_\odot$ in a time-scale of a few
$\tau_d$. Since $M_{d0}< M_{\rm disk}/f$, this may indicate that the
dust is externally provided in the galaxy. 

If the dust is not provided internally, the dust may be brought by
merged galaxies. That is the case for Centaurus~A \citep{esp12a}. The
misalignment between the disk and the galaxy's photometric axis
(section~\ref{sec:result}) is favorable to the merger scenario. Based on
a semi-analytic model, \citet{fuj00d} indicated that the fraction of cD
galaxies that have cold gas brought through galaxy mergers induced by
dynamical friction is small. However, their criterion for existence of
cold gas in a cD galaxy is $10^9\: M_\odot$, and thus more cD galaxies
should have cold gas of $<10^9\: M_\odot$ (see
\cite{tut07a}). Alternatively, a complicated solution may be
required. For example, while dust provided by evolved stars in a cD
galaxy is shielded from the surrounding hot gas \citep{don11a}, it is
mixed with cold gas provided by a weak cooling flow \citep{gas12a}.

\subsection{Comparison with Other cD Galaxies}

Among cD galaxies with outstanding dust lanes and disks, those of
Cygnus~A and Hydra~A are known for having extremely active radio sources
\citep{kin05a,nul05b}. The cD galaxy of Cygnus~A has a well-developed
cold disk rotating around the AGN and the minor axis of the disk is
parallel to the jets like Hydra~A \citep{jac98a,you02a}. On the other
hand, the cD galaxy of Hercules~A does not have a disk-structure,
although it has prominent twin jets \citep{ode13a}. This may be because
the cold disk has disappeared by now, while it caused the strong
outburst in the past. Note that in some non-cD radio galaxies, dust
lanes that resemble the one we found in Hydra~A have been observed
\citep{dek00a,der02a}.

\section{Conclusions}

We observed the cD galaxy of the Hydra~A cluster with Subaru telescope
and found an outstanding dust lane. The disk seen as the dust lane is
edge-on and the minor axis is parallel to radio jets, but it is not
parallel to the minor axis of the host galaxy. The disk mass and the
rotation period are $\sim 8\times 10^7\: M_\odot$ and $\sim 4\times
10^7$~yr, respectively. The column density of the dust lane ($N_{\rm
H}=2.0\times 10^{21}\rm\: cm^{-2}$) is consistent with the constraints
obtained by X-ray observations. Our results may indicate that galaxy
mergers and the supply of cold gas to the SMBHs is essential to the
most powerful activities of the AGNs.

\bigskip

This work was supported by KAKENHI (YF: 23540308, KS: 25800112), and
World Premier International Research Center Initiative (WPI Initiative),
MEXT, Japan


\end{document}